\documentclass[a4paper,prb,twocolumn,showpacs]{revtex4}
\usepackage{amsfonts}
\usepackage{graphicx}
\usepackage{color}
\usepackage{amsmath}
\usepackage{amssymb}

\begin{document}
\title{Model of the thermoelectric properties of anisotropic organic semiconductors}
\author{S. Ihnatsenka}
\affiliation{Department of Science and Technology, Link\"{o}ping University, SE-60174, Norrk\"{o}ping, Sweden}
\email{sergey.ignatenko@liu.se}

\begin{abstract}
A model of charge hopping transport that accounts for anisotropy of localized states and Coulomb interaction between charges is proposed. For the anisotropic localized states the degree of orientation relates exponentially to the ratio of conductivities in parallel and perpendicular directions, while the ratio of Seebeck coefficients stays nearly unaffected. However, the ratio of Seebeck coefficients increases if Coulomb interaction is screened stronger in a direction parallel to the predominant orientation of the localized states. This implies two different physical mechanisms responsible for the anisotropy of thermoelectric properties in the hopping regime: electronic state localization for conductivities, and screening for Seebeck coefficients. This provides explanation for recent experimental findings on tensile drawn and ribbed polymer films.
\end{abstract}
\maketitle

\section{Introduction}
Recently, rubbing and tensile drawing have been proposed as methods to enhance thermoelectric efficiency of conjugated polymers.\cite{Unt20, Ham17, Hyn19, Sch20, Vij19, Vij19_2} This type of mechanical processing makes otherwise random orientation of long polymer chains to become uniaxially aligned. A high degree of orientation has been directly observed by polarized optical microscopy,\cite{Unt20, Ham17, Vij19, Vij19_2} transmission electron microscopy\cite{Ham17, Vij19} and wide-angle X-ray scattering.\cite{Hyn19} For example, Untilova \textit{et al.}\cite{Unt20} found that Poly(3-hexylthiophene) (P3HT) doped with Mo(tfdCOCF$_3$)$_3$ reveals conductivity along the rubbing direction ($\sigma_{\parallel}$) that is 2.6 times larger than conductivity in the isotropic samples ($\sigma$) of the same compound, and by the absolute value $\sigma_{\parallel}$ exceeds all of the previous experimentally measured conductivities on P3HT. In contrast, conductivity $\sigma_{\bot}$ was by an order of magnitude smaller in the perpendicular direction: for oxidation level 11$\%$, $\sigma_{\parallel}=681$ S/cm, $\sigma_{\bot}=50$ S/cm, $\sigma=260$ S/cm.\cite{Unt20} Similar ratios and large absolute values along the rubbing direction have been observed for the Seebeck coefficient: $S_{\parallel}/S_{\bot}=7.1$ for oxidation level 11$\%$. In the experiment by Hynynen \textit{et al.},\cite{Hyn19} P3HT doped with Mo(tfdCOCF$_3$)$_3$ revealed strongly increased conductivity along the drawing direction, whereas the Seebeck coefficient was surprisingly unaffected. Apart from rubbing and tensile drawing, anisotropic P3HT films with similar properties have also been fabricated by a different technique, where fiber morphology was created via epitaxial growth and temperature-gradient crystallization when organic small-molecule 1,3,5-trichlorobenzene particles were added to the solution.\cite{Qu16}  Experimentally\cite{Unt20, Ham17, Hyn19, Sch20, Vij19, Vij19_2, Qu16} measured large conductivities make \textit{anisotropic} polymer films attractive as an electrode material in printed electronics,\cite{Ber07} while their large power factors make them attractive for application in thermoelectric generators.\cite{Bub12}

Theoretical treatment of enhanced thermoelectric properties of anisotropic polymer is limited to kinetic Monte-Carlo\cite{Sch20} and resistor network studies,\cite{Hyn19} in which several drawbacks are faced. For example, the latter approach is limited to 2D transport with positional dependence of the tunneling rates and interaction between charges both disregarded. Kinetic Monte-Carlo modeling in Ref. \onlinecite{Sch20} included Coulomb interaction between charges. However, only “on-site” interaction was taken into account, which neglects long-range nature of the repulsive Coulomb force between charged particles. Neglecting the long-ranged part of Coulomb interaction results in incorrect ground state and inability to capture the Coulomb gap, which is a fundamental property of the disordered system of the localized states\cite{Shk_book, Pol70, Efr75} that has been confirmed experimentally.\cite{Mas95, But00} The long-ranged part is particularly important for mediums with low dielectric permittivity to which organic semiconductors belong to. Typical permittivity of organic semiconductor is about 3.\cite{Bub12, Coe12, Pas05} Furthermore, both theories\cite{Sch20, Hyn19} use a simplified model of spatial anisotropy of the localized states and also use exponential density of states (DOS), which contradicts a number of studies, which have pointed out that Gaussian DOS is more accurate for disordered organic semiconductors.\cite{Bar14, Pas05, Oel12, Hul04, Yog11, Hes80} Both theories predict many-fold increase of $\sigma_{\parallel}/\sigma_{\bot}$ in agreement with experiments.\cite{Unt20, Ham17, Hyn19, Sch20, Vij19, Vij19_2, Qu16} That increase was attributed to increase of the anisotropy degree of the localized states. However, $S_{\parallel}/S_{\bot}$ disagreed with experimental data in Refs. \onlinecite{Unt20, Ham17, Sch20, Vij19, Vij19_2} where this ratio was 3-7 while the theory\cite{Sch20} predicted only 1.1-1.4 for the parameter range where agreement on $\sigma_{\parallel}/\sigma_{\bot}$ was achieved. This implies \textit{different} physical mechanism responsible for $S$ anisotropy. Thus, the origin of enhanced thermoelectric properties of anisotropic organic semiconductors observed in recent experiments\cite{Unt20, Ham17, Hyn19, Sch20, Vij19, Vij19_2} has remained an open question. 

A typical organic semiconductor contains a blend of conducting polymer molecules, dopants and insulating host molecules.\cite{Kai00, Kli06} A polymer is a long molecule that is composed of many repeating subunits (C$_4$H$_2$S for P3TH) that join together by covalent bonds. Due to covalent bonding, the electron wave function is delocalized along the molecule. Therefore, the whole molecule can be represented by a single localized state that is highly anisotropic in space. For an isotropic material, a blend of randomly oriented states gives zero net anisotropy, similar to magnetic dipoles in a paramagnet in the absence of a magnetic field. Rubbing or tensile drawing,\cite{Unt20, Ham17, Hyn19, Sch20, Vij19, Vij19_2} or the method proposed in Ref. \onlinecite{Qu16} --- in the first place --- change orientation and not the degree of anisotropy of the localized states (molecular structure of polymer). One of the aims of this study is to explore how orientation affects the thermoelectric properties of anisotropic materials. 

Another aim of this study is to understand the effects due to long-ranged Coulomb interaction on the thermoelectric properties of anisotropic materials, when unperturbed (single-electron) DOS is characterised by the Gaussian distribution. 

To achieve these objectives, a model of charge hopping transport is formulated that accounts for the effects of the orientation of anisotropic localized states and long-range Coulomb interactions between those states. The results that are presented in this manuscript show that $\sigma_{\parallel}/\sigma_{\bot}$ increases exponentially with a degree of orientation, while $S_{\parallel}/S_{\bot}$ remains nearly constant. Maximum power factors for making an efficient power generator can be attained at intermediate orientation degree, where less geometrical constraints apply on the current flow. Visualization of the current flow reveals predominant path along the host orientation. For typical material parameters and at room temperature, the charge transport in both parallel and perpendicular directions occurs as 3D hopping and at a crossover between variable range hopping (VRH) and nearest-neighbour hopping (NNH) regimes. The Coulomb interaction does not affect the orientational dependencies of $\sigma$ and $S$. However, in comparison to the non-interacting theory the ratio $S_{\parallel}/S_{\bot}$ can increase if the Coulomb interaction is screened more strongly in one direction than another, specifically in a direction parallel to the predominant orientation of the localized states. This anisotropic screening might be a result of the larger extent of the electron wave functions along polymer backbone chains when a system behaves like a metallic. Screening of the electric field inside anisotropic organic semiconductor, which depends on morphology as well as presence of conducting layers nearby (like metal gate electrode), explains why in some experiments\cite{Qu16, Hyn19} $S_{\parallel}/S_{\bot}\approx1$ but in the others\cite{Unt20, Ham17, Sch20, Vij19, Vij19_2} $S_{\parallel}/S_{\bot}> 1$, while $\sigma_{\parallel}/\sigma_{\bot}> 1$ in all of them. These results provide a microscopic explanation for the thermoelectric properties of anisotropic polymer films in recent experiments.\cite{Unt20, Ham17, Hyn19, Sch20, Vij19, Vij19_2, Qu16}

\section{Model}

\begin{figure}[t!]
\includegraphics[keepaspectratio,width=1.0\columnwidth]{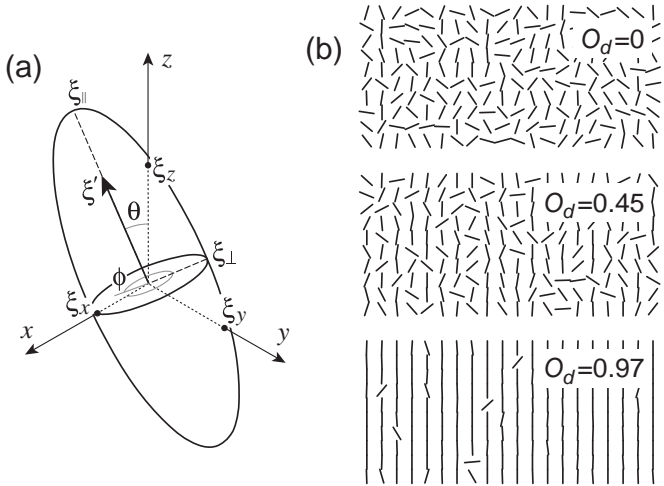}
\caption{(a) The localized state approximated by an ellipsoid of revolution. (b) Arrangement of the localized states for different degrees of orientation $O_d$.}
\label{fig:1}
\end{figure}

The hopping conduction between localized states in a disordered system is modeled by a resistor network.\cite{Mil60, McI79, Amb73} The resistance between two states $i$ and $j$ is\cite{Shk_book}
\begin{equation} \label{eq:2}
 R_{ij}=\frac{k_BT}{e^2\Gamma_{ij}},
\end{equation}
where the average tunneling rate accounting for wave function anisotropy is
\begin{widetext}
\begin{equation} \label{eq:3}
\Gamma_{ij}=\gamma_0\exp\left(-2\sqrt{\frac{x_{ij}^2}{\xi_{xi}\xi_{xj}}+\frac{y_{ij}^2}{\xi_{yi}\xi_{yj}}+\frac{z_{ij}^2}{\xi_{zi}\xi_{zj}}} -\frac{|E_i-E_j|+|E_i-\mu|+|E_j-\mu|}{2k_BT}\right),
\end{equation}
\end{widetext}
with $\gamma_0$ being the electron-phonon coupling parameter, ($x_{ij}$, $y_{ij}$, $z_{ij}$) are coordinate components of the vector connecting $i$ and $j$ sites, $E_i$ is the energy of the $i$-th state, $\mu$ is the chemical potential.

The localized electronic states are characterised by ellipsoids of revolution with semi-major and semi-minor axes, $\xi_{\parallel}$ and $\xi_{\bot}$, Fig. \ref{fig:1}(a). Ratio $\xi_{\parallel}/\xi_{\bot}$ describes the degree of anisotropy. Each ellipsoid is tilted with respect to the reference frame by a random pair of angles $\theta$ and $\varphi$. The unit vector $\xi^{\prime}$ defines the direction of $\xi_{\parallel}$, Fig. \ref{fig:1}(a). The Cartesian coordinates of the localized states in \eqref{eq:3} are the distances from the origin to the cross points of the ellipsoid surface with corresponding axes 
\begin{align} 
\xi_x &=\sqrt{\xi_{\parallel}^2 \cos^2\varphi +\xi_{\perp}^2(1-\cos^2\varphi)}, \\
\xi_y &=\sqrt{\xi_{\parallel}^2 \sin^2\varphi +\xi_{\perp}^2(1-\sin^2\varphi)}, \\
\xi_z &=\sqrt{\xi_{\parallel}^2 \cos^2\theta+\xi_{\perp}^2(1-\cos^2\theta)}.
\end{align}
Extra subscripts $i$ and $j$ in \eqref{eq:3} denote the lattice sites. Note that the minor principal axes are equal for an ellipsoid of revolution, but $\xi_x\neq \xi_y\neq\xi_z$ in general. For the isotropic state $\xi_{\parallel}=\xi_{\bot}=\xi_x=\xi_y=\xi_z$, and \eqref{eq:3} reduces to a familiar expression for the tunneling rate.\cite{Shk_book}

An average orientation of the localized states is represented by symmetric second-order tensor, which is calculated as the dyadic product of the unit vectors
\begin{equation*}
\mathrm{\mathbf{T}}=
\begin{bmatrix}
t_{xx} & t_{xy} & t_{xz} \\
t_{yx} & t_{yy} & t_{yz} \\
t_{zx} & t_{zy} & t_{zz} 
\end{bmatrix}
\end{equation*}
where
\begin{equation}
t_{\alpha\beta}=\frac{1}{N}\sum_{n=1}^{N} (\xi_{\alpha}^{\prime}\xi_{\beta}^{\prime})_n
\end{equation}
$\alpha,\beta=x,y,z$ ($\xi_x^{\prime}$ is $x$ component of the unit vector $\xi^{\prime}$) and the sum runs over all ellipsoids. The degree of orientation can be quantified by a scalar value $O_d$, which describes the strength of the main orientation of the tensor $\mathrm{\mathbf{T}}$ and is obtained from the largest eigenvalue of $\mathrm{\mathbf{T}}$. All of the eigenvalues are nomalized to unity, so the lowest possible value for the largest eigenvalue is $\frac{1}{3}$. The degree of orientation is thus conveniently written as
\begin{equation}
O_d=\frac{3}{2}\left( \lambda_1 - \frac{1}{3}\right)
\end{equation}
to make $O_d\in[0,1]$; $\lambda_1$ is the largest eigenvalue. The orientation of individual states is obtained from random distribution of the tilt angles of $\xi^{\prime}$, see Fig. \ref{fig:1}(b) for three representative orientations.

A network of the localized states is put on simple cubic lattice (3D Cartesian grid) with unit constant $l$. No positional disorder is assumed.

Here, two models for energies $E_i$ are considered: a non-interacting model and a model with Coulomb interaction between charged particles. In the non-interacting model, $E_i=E_i^0$ with $E_i^0$ randomly generated from the Gaussian distribution having the standard deviation $\sigma_{\textnormal{DOS}}$. 
In the interacting model, the energies are additionally renormalised leading to the local mean-field equations\cite{Ami08}
\begin{equation} \label{eq:ECoulomb}
E_i=E_i^0+\sum_{j\neq i}\frac{e^2}{r_{ij}}\left( \frac{1}{1+e^{\frac{E_j-\mu}{k_BT}}}-Q_b\right),
\end{equation}
where $r_{ij}$ is the distance between sites $i$ and $j$, $Q_b$ is the positive background charge equal to the relative charge concentration in the non-interacting model. The summation runs over all of the lattice indices and takes only the shortest distance between two sites in the repeated lattice; periodic boundary conditions are applied. Coulomb interaction results in the electrons moving in the average potential generated by all other electrons. This model is known\cite{Ami08} to correctly reproduce the Coulomb gap and Efros-Shklovskii VRH at low temperatures.\cite{Efr75, Shk_book} 

The method to find conductivity $\sigma$ and current densities $I$ is described in Ref. \onlinecite{Ihn16}. The system is assumed to be in a linear Ohmic regime. The chemical potential determines the charge density
\begin{equation}
n(\mu)=\frac{N_0}{\sqrt{2\pi}\sigma_{\textnormal{DOS}}}\int dE \exp\left(-\frac{E^2}{2\sigma_{\textnormal{DOS}}^2}\right) f(E,\mu),
\end{equation}
where $N_0=l^{-3}$ is the concentration of sites and $f$ is the Fermi-Dirac distribution function. The Seebeck coefficient, or thermopower, is given by
\begin{equation} \label{Seebeck}
S=-\frac{\pi^2 k_B^2 T}{3|e|}\frac{\partial}{\partial E} \ln\left[\sigma(E)\right]\vert_{E=\mu}.
\end{equation}
If conductivity is determined by diffusion and drift of non-interacting particles, then the Einstein relation can be applied to write
\begin{equation}
\sigma=e^2\rho D,
\end{equation}
where $\rho$ is DOS at the chemical potential and $D$ is the diffusion coefficient. The Seebeck coefficient \eqref{Seebeck} thus becomes
\begin{equation} \label{Seebeck2}
S\propto \frac{1}{\rho}\frac{d\rho}{dE}+ \rho \frac{d(\ln D)}{dn}.
\end{equation}
Because $D$ depends weakly on $n$, $S$ is mostly determined by $\rho$ and its slope over energy.

\section{Results and discussion}

The numerical calculations are performed for a parameter set that is typical for organic semiconductors.\cite{Bar14, Pas05, Oel12, Kim12_Pipe, Coe12, Ihn15} In particular, $\xi_{\parallel}$ is chosen to be equal to the lattice constant, a value large enough not to bring the system into a strong localization (insulating) regime. The anisotropy of the localized states is $\xi_{\parallel}/\xi_{\bot}=4$. The lattice constant $l=1$ nm. The electron-phonon coupling  $\gamma_0=10^{13}$ s$^{-1}$. The strength of energetic disorder is $\sigma_{\textnormal{DOS}}=0.1$ eV. $T=300$ K. The disorder is assumed to be only energetic and orientational; the effect of positional disorder will be commented on later. The system size for the results presented below is $20 \times 20 \times 20$, unless otherwise stated. Averaging is performed over 100 different disorder realizations. The calculations were also performed for different sizes, $\xi_{\parallel}$ and $\xi_{\bot}$, and similar results were obtained. 

To understand how the orientation of the localized states affects the thermoelectric properties, let us first consider the noninteracting theory. 

\begin{figure}[h]
\includegraphics[keepaspectratio,width=\columnwidth]{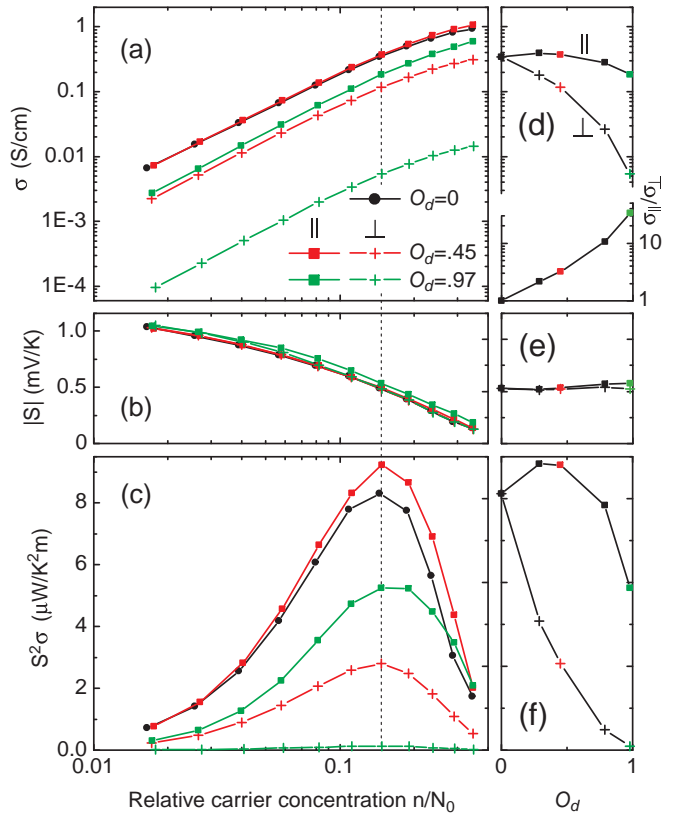}
\caption{The conductivity $\sigma$, Seebeck coefficient $S$ and power factor $S^2\sigma$ as a function of the relative charge concentration (a)-(c) and orientation degree (d)-(f). The vertical-dotted line denotes $n$ for optimal doping level, which is fixed in (d)-(f) so the right panels show the evolution of $\sigma$, $S$ and $S^2\sigma$ with $O_d$ for the optimal doping level. The lower part of (d) shows ratio $\sigma_{\parallel}/\sigma_{\bot}$. $\xi_{\parallel}/\xi_{\bot}=4$. }
\label{fig:2}
\end{figure}

Conductivity, Seebeck coefficient and power factor $S^2\sigma$ (PF) for a system of anisotropic localized states follow similar concentration dependence to that of isotropic states,\cite{Kim12_Pipe, Bub11, Ihn15} Figs. \ref{fig:2}(a)-(c). As charge concentration (or chemical potential) increases, the effects on $\sigma$ and $S$ go in opposite directions, so the maximum of PF occurs at some intermediate $n$ that is referred to as an optimal doping level.\cite{Bub11} This is a desirable value for making an efficient thermoelectric generator.\cite{Bub12} $n$ is directly proportional to the oxidation level measured in the experiments.\cite{Kim12_Pipe, Ihn15} In the absence of a predominant orientation, the network of anisotropic localized states is a system with random spatial extents of the wave functions localized on the lattice sites and it acts as if it is made of the isotropic states but randomly distributed in position. The macroscopic quantities, such as $\sigma$, are thus direction independent. \textit{When uniaxial orientation starts to reveal, it clearly manifests itself in $\sigma_{\parallel}$ and $\sigma_{\bot}$, whose ratio exponentially increases with $O_d$}, Figs. \ref{fig:2}(a),(d). For moderate $O_d$, $\sigma_{\parallel}$ reveals a slight increase, which implies a more effective percolation for charges hopping through disordered medium, and $\sigma_{\parallel}>\sigma>\sigma_{\bot}$ similarly to the experimental data on P3HT in Ref. \onlinecite{Unt20}. Overall, however, $\sigma$ decreases with $O_d$ because the orientation of the localized states imposes a geometrical limitation on the conduction path. In the limiting case $O_d=1$, the network breaks down into a series of parallel connected 1D chains, part of which is blocked by strong potential fluctuations. \textit{Concentration dependence of $S$, in contrast, occurs being independent on $O_d$}, see Figs. \ref{fig:2}(b)(e) where negligibly small $S_{\parallel}/S_{\bot}=1.1$ develops for  $O_d=0.97$. This can already be understood from simpler considerations using \eqref{Seebeck2}: $S$ is proportional to DOS and its derivative over energy and because DOS follows the same Gaussian distribution irrespective of transport directions, $S$ does not depend on $O_d$. The power factor mainly reflects the dependence of $\sigma$ on $O_d$, Figs. \ref{fig:2}(c),(f), which is non-monotonic parallel to alignment direction. (Note that a log scale is used in (d) and a linear scale is used in (f).)  

\begin{figure}[h]
\includegraphics[keepaspectratio,width=\columnwidth]{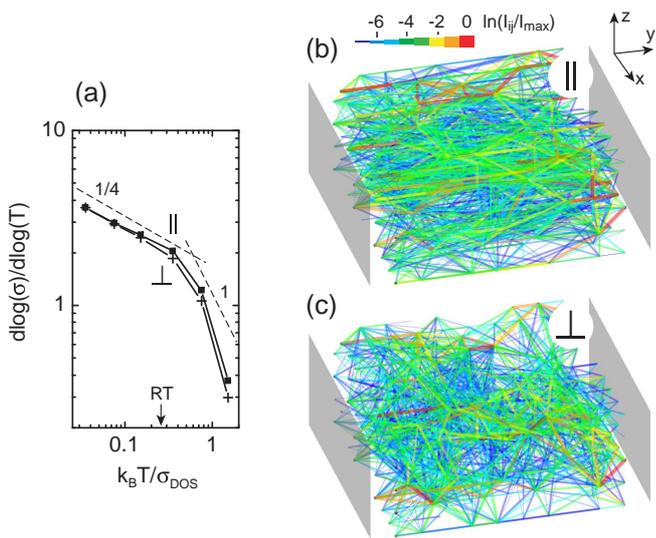}
\caption{Temperature dependence of reduced activation energy (a) and current density (b),(c) for anisotropic system with $\xi_{\parallel}/\xi_{\bot}=4$ and $O_d=0.45$. The dashed lines show the slopes for 3D VRH and NNH, exponents $\alpha=\frac{1}{4}$ and 1 in the Mott’s law $\sigma\propto\exp (T_0/T)^{\alpha}$, where $T_0$ is a characteristic temperature.\cite{Mot69} In (b) and (c), the dots mark the hopping sites with size inversely proportional to the absolute energy difference to $\mu$; same energetic disorder in (b) and (c). The gray pads are the source and drain electrodes. $T$ is chosen to be room temperature (RT in (a)), which for a typical organic semiconductor\cite{Bar14, Pas05, Oel12, Kim12_Pipe, Coe12, Ihn15} relates to the degree of disorder as $4k_BT=\sigma_{\textnormal{DOS}}$. Lattice $10\times 10\times 5$ is shown. $n/N_0=0.1$. }
\label{fig:4}
\end{figure}

The transport regime can be determined from the temperature dependence of the reduced activation energy,\cite{Zab84} Fig. \ref{fig:4}(a), which suggests that for typical values of $\sigma_{\textnormal{DOS}}$,\cite{Bar14, Pas05, Oel12, Kim12_Pipe, Coe12, Ihn15} the charge transport at room temperature occurs at a crossover between VRH and NNH. For a charge carrier, the phonon energy becomes insufficient to assist hopping to the nearest localized states and hopping to the distant states becomes energetically favourable. This can be directly observed in the current density visualization, both parallel and perpendicular to the alignment direction, see Figs. \ref{fig:4}(b) and (c). The charge flow spans uniformly over device volume and reveals a substantial degree of anisotropy that reflects the underlying orientation of the localized states and their anisotropy. If the dominant orientation is perpendicular to the current flow direction, see Fig. \ref{fig:4}(c), then the electrons adjust their path further in the perpendicular direction to find a site to hop in that is closer in energy.\cite{Ihn16}  The average hopping length might be estimated from the length of individual current segments, which is $2.12 l$ and $2.07 l$ in Figs. \ref{fig:4}(b) and (c), respectively.

The results of the noninteracting theory show that the uniaxial orientation of the anisotropic states results in an exponential increase of the ratio $\sigma_{\parallel}/\sigma_{\bot}$, while $S_{\parallel}/S_{\bot}$ stays nearly constant and the transport regime is at the crossover of VRH and NNH. Similar ratios were observed in the experiments on P3HT films in Refs. \onlinecite{Qu16, Hyn19}. However, other experimental data on rubbed and tensile drawn polymer films,\cite{Unt20, Ham17, Sch20, Vij19, Vij19_2} including P3HT,\cite{Unt20, Ham17, Vij19} revealed a simultaneous manifold increase of $\sigma_{\parallel}/\sigma_{\bot}$ and $S_{\parallel}/S_{\bot}$ in the highly oriented state. One of the reasons that might cause the latter observations is the effect of electron-electron interaction, which was not been taken into account in the above results.

\begin{figure}[h]
\includegraphics[keepaspectratio,width=0.775\columnwidth]{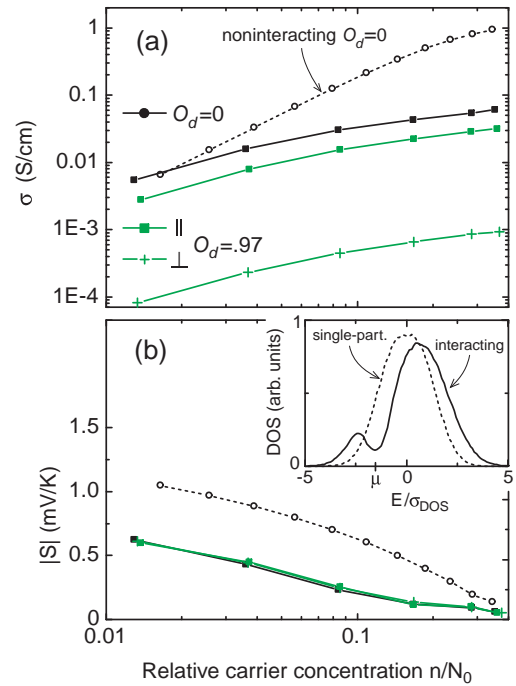}
\caption{(a) The conductivity $\sigma$ and (b) Seebeck coefficient $S$ as a function of the relative charge concentration in the theory with Coulomb interaction taken into account. The dotted lines with open circles correspond to $O_d=0$ in the noninteracting theory and are given for comparison from Figs. \ref{fig:2}(a),(b). The dependence on $O_d$ is the same as in the non-interacting theory; only $O_d=0.97$ is shown. The inset in (b) compares the single-particle (dotted line) and interacting (solid line) DOS at $\mu/\sigma_{\textnormal{DOS}} =-1.5$. $\xi_{\parallel}/\xi_{\bot}$=4.}
\label{fig:5}
\end{figure}

Figure \ref{fig:5} shows how Coulomb interaction affects concentration dependence of $\sigma$ and $S$. For the same parameter set as in the noninteracting modeling the concentration dependence of $\sigma$ occurs more weakly. This is a result of DOS renormalization due to Coulomb repulsion between charge carriers; see inset in Fig. \ref{fig:5}(b), where the Coulomb gap\cite{Pol70, Efr75, Shk_book} at $\mu$ is clearly seen. Note that existence of the Coulomb gap in the DOS spectrum of disordered systems was confirmed in the electron tunneling experiments.\cite{Mas95, But00} At high concentrations, the number of states available for conduction is reduced and thus $\sigma$ becomes smaller when compared to the noninteracting case. At $n/N_0\approx 0.015$, the effect due to electron interaction on $\sigma$ diminishes and reverts at lower $n$. The DOS shape at the Coulomb gap is close to symmetric, even though the single-particle $\rho$ is the exponential function of energy, which results in smaller absolute values of the Seebeck coefficient, as can already be obtained from Eq. \eqref{Seebeck2}. Orientational dependence is not affected by Coulomb interaction because the renormalized energies \eqref{eq:ECoulomb} are scalars entering the tunneling rates (the second term in \eqref{eq:3}). For the sake of visual clarity, only $O_d=0$ and $O_d=0.97$ are presented in Fig. \ref{fig:5}.

While the interacting theory alone cannot explain the findings on $S_{\parallel}/S_{\bot}$ in Refs. \onlinecite{Unt20, Ham17, Sch20, Vij19, Vij19_2}, an important result is that \textit{the Coulomb interaction causes a many-fold suppression of the Seebeck coefficient in comparison to the non-interacting theory for the same system}. Thus, experimental findings\cite{Unt20, Ham17, Sch20, Vij19, Vij19_2} might be explained to the dominance of the Coulomb interaction in only one (perpendicular) direction. This can in turn be rationalized by the fact that in the parallel direction of the highly-ordered organic semiconductor, the wave functions overlap strongly and screening of the electric filed is more effective. Note that the mean-field theory cannot capture this effect because it is formulated for point-like charges that interact via unscreened (isotropic) Coulomb potential. Screening might be introduced into the theory via the dielectric constant $\kappa$ by replacing $e^2$ by $\frac{e^2}{\kappa}$ in \eqref{eq:ECoulomb} and further assuming that it is direction dependent. (Examples of materials with anisotropic $\kappa$ include barium titanate, black phosphorus and nematic liquid crystals.) In the case of perfect screening, $\kappa_{\parallel}\rightarrow\infty$ and $S_{\parallel}$ is given by the non-interacting theory. If $\kappa_{\bot}=1$, $S_{\bot}$ is given by the result from the interacting theory. Therefore, $S_{\parallel}/S_{\bot}$ would be the ratio between the values in the noninteracting and interacting theories and, for $n/N_0=0.1$, $S_{\parallel}/S_{\bot}\approx2.5$ in Fig. \ref{fig:5}(b). This estimate should be lower for the organic semiconductors which have $\kappa\approx 3$.\cite{Bub12, Coe12, Pas05} To increase $S_{\parallel}/S_{\bot}$ the anisotropy of the localized states $\xi_{\parallel}/\xi_{\bot}$ should increase provided by the condition $O_d\rightarrow1$. For every experimental sample in Refs. \onlinecite{Unt20, Ham17, Sch20, Vij19, Vij19_2}, the ratio $S_{\parallel}/S_{\bot}$ thus signifies strength of Coulomb interaction and screening abilities that are anisotropic in space. For the samples in Refs. \onlinecite{Qu16, Hyn19}, where $S_{\parallel}/S_{\bot}\approx 1$, it might be argued that electron-electron interactions are suppressed due to, for example, a nearby gate electrode or another conducting layer. It is then straightforward to implement a verification of this theory: Place a metal near the sample and check whether $S_{\parallel}/S_{\bot}$ decreases or not.

Transport regimes obtained in the noninteracting model should still be valid for the experimental conditions\cite{Unt20, Ham17, Sch20, Vij19, Vij19_2, Qu16, Hyn19} even though electron interactions are included into theory because Efros-Schklovskii VRH occurs at much lower $T$.\cite{MottES} The crossover between Mott's VRH and NNH depends primarily on wave function localization, which can be demonstrated using Mott’s hopping theory.\cite{Mot69} In the VRH regime, the average hopping distance is\cite{Shk_book}
\begin{equation}
r=\left(\rho \epsilon_0\right)^{-\frac{1}{3}},
\end{equation}
where $\rho$ is DOS at the chemical potential and
\begin{equation}
\epsilon_0=\frac{(k_BT)^{-\frac{3}{4}}}{(\rho \xi^3)^{-\frac{1}{4}}}
\end{equation}
is the energy at the percolation threshold. At the crossover $r\approx l$. Thus, the following relation holds
\begin{equation} \label{eq:crossover}
k_BT=\frac{\xi}{l^4 \rho},
\end{equation}
which implies that as the wave functions become more localized on the trapping sites, a lower temperature is required to bring the system from NNH into VRH regime. The relation \eqref{eq:crossover} was obtained for $\rho$ being constant over at least the scale of $k_BT$. This is not fulfilled for Gaussian distribution, but the analysis similar to Ref. \onlinecite{Zvy08} can be applied to show that this relation should also be valid for Gaussian DOS. Note also that the Mott’s law with exponent $\frac{1}{4}$ holds for Gaussian DOS in Fig. \ref{fig:4}(a). The VRH-NNH crossover depends on wave function localization but not on the degree of orientation $O_d$. This seemingly contradicts the argument in Ref. \onlinecite{Vij19} that different transport mechanisms exist in parallel and perpendicular directions. In Fig. \ref{fig:4}, VHR regime corresponds to 3D transport because the parameters of the anisotropic system still allow significant hopping rates in the transverse direction. For 1D VRH to take place, the degree of anisotropy should be much stronger.

The orientational dependence of PF in Fig. \ref{fig:2}(f) implies that a more effective power generator can be built by aligning the localized states to some moderate degree but not to the full extent. This might naturally be the case in the experimental samples due to misalignment at the boundaries between crystalline grains.\cite{Kli06} It is interesting that an ideal single crystal for organic thermoelectrics (in the hopping regime) is not the solution since it would reduce PF.  An experiment of gradual uniaxial alignment, including much higher degrees of alignment and concentration measurements, would be of interest to verify the results presented here.

The theoretical results presented above agree qualitatively with the experimental data\cite{Unt20, Ham17, Sch20, Vij19, Vij19_2, Qu16, Hyn19} for the ratios $\sigma_{\parallel}/\sigma_{\bot}$ and $S_{\parallel}/S_{\bot}$. However, absolute values of $\sigma$ and $S$ differ; in particular, $\sigma$ is a few orders of magnitude smaller. This can be explained by the usage of typical parameters\cite{Bar14, Pas05, Oel12, Kim12_Pipe, Coe12, Ihn15} for modeling without any additional tweaking. Quantitative agreement for $\sigma$ and $S$ is left for future study. Note that experimental data\cite{Unt20, Ham17, Hyn19, Sch20, Vij19, Vij19_2, Qu16} largely vary from one sample to another because of the strong sensibility of morphology to the preparation process. For P3HT the resulting polymer film might generally be amorphous, crystalline or a mix of the two.\cite{Kli06} Thus quantitative estimation of the parameters entering the theory should be done for every sample individually.

Apart from the effects introduced by Coulomb interaction in comparison to the non-interacting theory as shown above it is instructive to consider the consequences of truncating the long-ranged part of the electron-electron interaction. This can be done using the classic argument for the existence of the Coulomb gap in the distribution of energy levels of localized electrons in strongly disordered semiconductors.\cite{Efr75} Consider a pair of states $i$ and $j$ respectively above and below $\mu$. The stability criteria for the ground state requires that\cite{Shk_book}
\begin{equation} \label{eq:ground_state}
E_j-E_i>\Delta_{ij}\simeq \frac{e^2}{r_{ij}},
\end{equation}
where $\Delta_{ij}$ is positive and signifies that the energy of the ground state cannot be lowered by promoting an electron from $j$ to $i$. The states on opposite sides of $\mu$ that differ in energy by less than a small value $\eta$ must be separated in space by a distance larger than $e^2/\eta$. Hence, the spatial DOS vanished at least as fast as $(\eta/e^2)^d$, where $d$ is the dimensionality of the system. If long distances are removed from consideration the smallness of $\eta$ is never achieved and DOS does not vanish at $\mu$. 

Several final comments follow. First, if positional disorder is added to the modeling, the orientational dependence does not change appreciably. Second, both non-interacting and interacting models predict $S$-$\sigma$ dependence to have a fall-off shape, similarly to other hopping theories,\cite{Sch20, Unt20} but in contrast to experimentally observed trend $S\propto \sigma^{-1/4}$.\cite{Gla15} That was shown to be due a limitation of the VRH model itself.\cite{Kan17} Third, the hopping rates \eqref{eq:3} assume electrons or holes as charge carriers. These rates are modified when polaron effects become important. Those effects, however, are expected to be small for the system parameters and the linear Ohmic regime studied here.\cite{Small_polaron}

\section{Conclusion}
A charge hopping model is presented that accounts for the correlation between deformational and orientational degrees of freedom of the localized states, energetic disorder, electron-electron interactions and charge concentration. For a parameter set similar to typical organic semiconductors,\cite{Bar14, Pas05, Oel12, Kim12_Pipe, Coe12, Ihn15} it is shown that an increase of the degree of orientation causes an exponential increase of the ratio of conductivities in parallel and perpendicular directions, while the ratio of Seebeck coefficients stays nearly unaffected. However, the ratio of Seebeck coefficients can increase if Coulomb interaction is taken into consideration and charge screening in the direction parallel to the predominant orientation of the localized states is stronger than in the perpendicular direction. The regime of charge transport occurs at the crossover between VRH and NNH in both directions. These findings provide a microscopic explanation for the thermoelectric properties of anisotropic polymer films in recent experiments\cite{Unt20, Ham17, Hyn19, Sch20, Vij19, Vij19_2, Qu16} and show how those properties can be further tailored.


\section{Acknowledgement}
This work was supported by SNIC 2020/13-95. It is a pleasure to acknowledge discussion with X. Crispin.

\end{document}